\documentclass[amsmath,amssymb,twocolumn,nofootinbib]{revtex4-1}
 \newcommand{\be}[1]{\begin{equation}\label{#1}}
 \newcommand{\ba}[1]{\begin{eqnarray}\label{#1}}
 \newcommand{\ep}[1]{\epsilon_{#1}}

 \newcommand{\rd}{{\rm d}}
 
 \newcommand{\re}{{\rm e}}
 \newcommand{\pa}[1]{\left(#1\right)}
 \newcommand{\paq}[1]{\left[#1\right]}
 
 \newcommand{\av}[1]{\langle#1\rangle}
 \newcommand{\M}{{\rm M_{\rm P}}}
 \newcommand{\Mt}{{\rm \widetilde M_{\rm P}}}

 \def\ee{\end{equation}}
 \def\ea{\end{eqnarray}}

\usepackage{graphicx,latexsym}
\usepackage{graphicx,epsf}
\usepackage{color}
\usepackage{epsfig}
\usepackage{amsmath}
\usepackage{amssymb}
\usepackage[T1]{fontenc}
\usepackage[utf8]{inputenc}
\usepackage{tikz}
\usepackage{pgfplots}
\begin{document}
\title{Quantum Gravity, Time, Bounces and Matter}
\author{Alexander~Yu.~Kamenshchik}
\email{kamenshchik@bo.infn.it}
\affiliation{Dipartimento di Fisica e Astronomia, Universit\`a di Bologna\\ and INFN,  Via Irnerio 46, 40126 Bologna,
Italy,\\
L.D. Landau Institute for Theoretical Physics of the Russian
Academy of Sciences,\\
 Kosygin street 2, 119334 Moscow, Russia}
 \author{Alessandro~Tronconi}
\email{tronconi@bo.infn.it}
\affiliation{Dipartimento di Fisica e Astronomia, Universit\`a di Bologna\\ and INFN, Via Irnerio 46, 40126 Bologna,
Italy}
\author{Tereza Vardanyan}
\email{tereza.vardanyan@bo.infn.it}
\affiliation{Dipartimento di Fisica e Astronomia, Universit\`a di Bologna\\ and INFN, Via Irnerio 46, 40126 Bologna,
Italy}
\author{Giovanni~Venturi}
\email{giovanni.venturi@bo.infn.it}
\affiliation{Dipartimento di Fisica e Astronomia, Universit\`a di Bologna\\ and INFN,  Via Irnerio 46, 40126 Bologna,
Italy}

\begin{abstract}
In the context of Quantum Cosmology and the Wheeler-DeWitt equation we investigate the possible effects of a non semiclassical wave-function of the universe on the evolution of the inflationary perturbations. These are associated with the quantum behaviour of the homogenous degrees of freedom (in particular the radius of the universe) in the early stages of the inflationary expansion, which in turn can affect the dynamics of the trans-Planckian modes of the fields present. The existence of a bounce for the homogeneous gravitational wave-function is studied. This can lead to an interference between a contracting and an expanding universe and, as a consequence, to the above quantum gravitational effects on the primordial spectra. In the traditional study of the inflationary fluctuations such effects are neglected and a quasi-classical behaviour for the homogeneous inflaton-gravity system is taken.
 \end{abstract}

\maketitle
\section{Introduction}
Thanks to numerous cosmological observations we have fairly precise information on the distribution of matter in the universe \cite{cmb}. Because of the inflationary paradigm \cite{inflation} this is related to quantum fluctuations at the beginning of the evolution of the universe when it was very small and presumably quantum effects were very important \cite{pert}. This has led to the study of the quantum matter-gravity system with the aim of understanding how time and the structures in the universe emerged. We have previously examined the matter-gravity quantum system in the context of a Wheeler-DeWitt (WdW) equation \cite{wdw}, quantum matter and a Born-Oppenheimer (BO) approach \cite{BO,BO-cosm0,BO-cosm} wherein gravitation is associated with the heavy (Planck mass) degrees of freedom and matter corresponds to the light ones. The introduction of the semiclassical limit for gravitation leads to the emergence of time and an evolution equation for matter (cosmological perturbations) having corrections involving the Planck mass.\\
An alternative approach we examined was the emergence of a time even with gravity and the inflaton in quantum regimes \cite{TVV}: it is a consequence of the fact that normal matter cannot see quantum oscillations above the Planck frequency but just experiences an evolution with respect to a function of the scale factor, associated with the speed of inflation. In such a framework time only exists for normal matter which evolves according to its position on the gravitational wave-function of the universe.
In particular such an approach consisted of the study of a quantum matter-gravity system containing a minimally coupled massive homogeneous scalar field which is known to lead to inflation. After choosing a suitable highly peaked initial state for the scalar field, the equation for the homogeneous gravity-matter system is solved in the inflationary (scale factor $a$ large) limit. On then introducing other matter fields (or inhomogeneous modes), after coarse graining of the gravitational wave-function, an effective time evolution emerges for them. In this case the presence of an effective time evolution for matter arises from a mechanism similar to one already observed in the analysis of the classical limit of quantum systems, such as the hydrogen atom \cite{rowe}. In particular, the zero angular momentum and large principal quantum number case which exhibits a radial highly oscillatory behaviour. In this case on coarse graining (in particular on applying the Riemann-Lebesgue Lemma) one is able to recover the classical trajectory. Indeed, the classical trajectory is related to a classical spatial probability distribution of a particle in terms of the inverse of its speed (the fraction of time spent in a spatial interval is a measure of the probability density \cite{Sommerfeld}). 

There is a deep connection between the above example and the situation present in the matter-gravity system for this case.\\
In recent years, there has been increasing interest in cosmological models that replace the initial cosmological singularity (or ``big bang'') with a ``big bounce'', i.e. a ``smooth'' transition from contraction to expansion. Many models of quantum gravity suggest the possibility of a bouncing Universe, since a point-like singularity is incompatible with the laws of quantum mechanics. Actually some hints of a bouncing dynamics were present in \cite{TVV} where the gravitational wave-function, in the large $a$ limit ($a$ being the scale factor), was found to be an arbitrary superposition of a contracting and an expanding homogeneous universe. Such an arbitrariness was due to the lack of an exact solution (or at least an approximate solution) for all $a$ and its effects on the evolution of normal matter (inflaton/metric fluctuations) were eliminated by the coarse graining. Averaging the dynamical equations for ``normal'' matter over Planckian oscillations is reasonable at energies below the Planck scale. However, during the inflationary expansion, the wavelength of inflaton/metric fluctuations is stretched from trans-Planckian values down to sub-Planckian ones and throughout their entire evolution a coarse graining procedure may not be justified any longer. \\
In this paper we calculate the quantum gravitational effects originating from the trans-Planckian oscillations of the gravitational wave-function for $a$ large on the inflationary spectra. That is we study the effects of an essentially quantum mechanical gravitational wave-function on the dynamics of the inflationary perturbations. Such effects arise from the superposition of two quantum states far from the classical regime and such a superposition can be justified by the existence of a bounce solution for the homogeneous WdW equation  in the $a$ small regime.\\
The paper is organised as follows: in Section 2 we introduce the general formalism for the classical and quantum description of inflation with a minimally coupled scalar field plus inflaton/metric perturbations. In  Section 3 we perform a BO decomposition for the inhomogeneous matter-gravity system. In  Section 4 we approximately solve the homogeneous WdW equation both for $a$ large (during inflation) and $a$ small and we justify the existence of a bouncing universe. In  Section 5 we study the quantum gravitational effects on the evolution of the inflationary perturbations and in Section 6 we apply our results to the de Sitter case. Finally  Section 7 is dedicated to the conclusions.

\section{Formalism}
We consider the inflaton-gravity system which is described by the following action
\be{fullaction}
S=\int d\eta d^3x\sqrt{-g}\paq{-\frac{\M^2}{2}R+\frac{1}{2}\partial_\mu\phi\partial^\mu\phi-V(\phi)}
\ee
where $\M=\pa{8\pi G}^{-1/2}$ is the reduced Planck mass.
The above action can be decomposed into a homogeneous part plus fluctuations around it. The homogeneous part determines the overall expansion while the inhomogeneous contributions are treated as perturbations. These perturbations play a major role as their spectra can be connected both to the dynamics of the homogeneous part and to the observable features of the CMB \cite{pert}. 
The fluctuations of the metric $\delta g_{\mu\nu}(\vec x,\eta)$ are defined by
\be{metricsc}
g_{\mu\nu}=g_{\mu\nu}^{(0)}+\delta g_{\mu\nu}
\ee 
where $g_{\mu\nu}^{(0)}=\rm{diag}\paq{a(\eta)^2\pa{1,-1,-1,-1}}$ is a flat RW metric and $\eta$ is the conformal time. Only the scalar and the tensor fluctuations ``survive'' the inflationary expansion: $\delta g=\delta g^{(S)}+\delta g^{(T)}$. The scalar inflaton fluctuation is defined as $\phi(\vec x,\eta)\equiv\phi(\eta)^{(0)}+\delta \phi(\vec x,\eta)$ and mixes with the scalar metric degrees of freedom $\delta g^{(S)}$. The physical perturbations can be finally described by three Mukhanov-Sasaki (MS) fields (one for the scalar part and two for the independent tensor polarisations). The homogeneous part plus the linearised perturbations dynamics is given by the following action
\ba{act}
S&=&\int d\eta\left\{L^3\paq{-\frac{\Mt^2}{2}a'^2+\frac{a^2}{2}\pa{\phi_0'^2-2V(\phi_{0})a^2}}\right.\nonumber\\
&+&\left.\frac{1}{2}\sum_{i=1,2}\sum_{k\neq 0}\paq{v_{i,k}'(\eta)^2+\pa{-k^2+\frac{z''}{z}}v_{i,k}(\eta)^2}\right.\nonumber\\
&+&\left.\frac{1}{2}\sum_{\lambda=+,\times}\sum_{i=1,2}\sum_{k\neq 0}\paq{\pa{\frac{v_{i,k}^{(\lambda)}}{d\eta}}^2+\pa{ -k^2 +\frac{a''}{a}}\pa{v_{i,k}^{(\lambda)}}^2 }\right\}\nonumber\\
&\equiv& S_G+S_I+S_{MS}
\ea
where the $v_{i,k}$ are Fourier components of the scalar MS field and the $v_{i,k}^{(\lambda)}$ are those of the MS tensor field and the index $i$ accounts for the real and imaginary parts of each component, $\Mt=\sqrt{6}\M$, $z\equiv \phi_0'/H$, $H=a'/a^2$ is the Hubble parameter and $L^3\equiv \int d^3x$. We formally split the full action into three contributions: $S_G$ and $S_I$ are the homogeneous gravity and inflaton actions respectively and $S_{MS}$ collectively describes perturbations.\\
Let us note that on working in a flat 3-space and considering both homogeneous and inhomogeneous quantities one must introduce an unspecified length $L$. One can then eliminate the factor $L^3$ by replacing $a\rightarrow a/L$, $\eta\rightarrow \eta L$, $v\rightarrow \sqrt{L} v$ and $k\rightarrow k/L$. Such a redefinition is equivalent to setting $L=1$ in the above action (\ref{act}) (then implicitly assuming the convention $[a(\eta)]=l$ and $[dx]=[d\eta]=l^0$) and then proceeding with its quantisation. Henceforth we shall use this latter simplifying choice. Only at the end, in order to compare our results with observations, we shall restore all quantities to their original definitions and the dependence on $L$ will become explicit. 
Let us finally note that the fact that $L$ is infinite does not create a problem. As usual, the transition from the Fourier integral w.r.t. the wave number to the Fourier series eliminates the corresponding divergence.\\
The action (\ref{act}) is the starting point for the study of the inflationary dynamics in the semiclassical context. In such a context one assumes that the homogeneous quantities have a classical behaviour and only the perturbations are quantised. Within such a framework the space-time coordinates are well defined classical labels and the scale factor and the homogeneous inflaton have a definite time dependence which is given by the classical Einstein equations. However in a quantum universe the above assumption is not true and the homogeneous d.o.f. must also be quantised. Furthermore the definition of space-time coordinates loses its classical (intuitive) meaning. The exact treatment of homogeneous plus inhomogeneous d.o.f. in a fully quantised framework suffers from many technical problems. This is the main reason behind either the minisuperspace approximation (which neglects the perturbations) leading to the homogeneous WdW equation or the semiclassical approximation for the inflatonic scalar-tensor fluctuations which are commonly adopted for (\ref{act}). If one needs to estimate the effects arising in a fully quantised system on the dynamics of the primordial fluctuations then these effects must be treated in a perturbative fashion. With this assumption the classical definition of space and time holds at zero order and one is finally led to the standard MS equations plus a perturbation which is a consequence of a fully quantised system. Moreover in the leading order one recovers the classical trajectory $a(\eta)$ and $\phi_0(\eta)$.
Technically this amounts to canonically quantising the homogeneous and the inhomogeneous parts separately, i.e. in treating the homogeneous variables which appear in $S_{MS}$ as classical (zero-order) time variables.\\
If one, without loss of generality, considers just one MS field then (\ref{act}) reduces to 
\ba{actS}
S&=&\int d\eta\left\{\paq{-\frac{\Mt^2}{2}a'^2+\frac{a^2}{2}\pa{\phi_0'^2-2V(\phi_{0})a^2}}\right.\nonumber\\
&+&\left.\frac{1}{2}\sum_{k\neq 0}^\infty\paq{v_{k}'(\eta)^2-\omega_k^2v_{k}(\eta)^2}\right\}\equiv\int d\eta \mathcal{L}_{tot}
\ea
where $\omega_k^2=k^2+m^2(\eta)$ is time dependent and $L$ has been set equal to $1$. Let us note that the time dependent mass in $\omega_k^2$ is $m^{2}(\eta)=-\frac{z''}{z}$ for each mode of the scalar perturbation and $m^{2}(\eta)=-\frac{a''}{a}$ for each mode of the tensor perturbation, where $z(\eta)$, $a(\eta)$ are classical (zero order) expressions\footnote{in particular $z=a\sqrt{\ep{1}}$ where $\ep{1}$ is the first Hubble flow function associated with slow roll;}. In our approach $\omega_k^2$ is finally rewritten as a function of $a$ and related to $\eta$ on the classical trajectory.\\
The Hamiltonian  is finally
\begin{eqnarray}
&&\mathcal{H}=-\frac{\pi_a^2}{2\Mt^2}+\frac{\pi_\phi^2}{2a^2}+a^4 V+\sum_{k\neq 0}^{\infty}\paq{\frac{\pi_k^2}{2}+\frac{\omega_k^2}{2}v_k^2}\nonumber \\
&&\equiv \mathcal{H}_G+\mathcal{H}_I+\mathcal{H}_{MS},
\label{ham}
\end{eqnarray}
where henceforth we shall limit ourselves to the chaotic inflaton potential $V=m^2\phi_0^2/2$ and
\be{momenta}
\pi_a=-\Mt^2 a'\,,\;\pi_\phi=a^2\phi_0'\,,\;\pi_k=v'_k.
\ee
The canonical quantisation of the Hamiltonian constraints leads to the following WdW equation for the wave-function of the universe (matter plus gravity)
\begin{widetext}
\be{ WdW0}
\left\{\frac{1}{2\Mt ^2}\frac{\partial^2}{\partial a^2}-\frac{1}{2a^2}\frac{\partial^2}{\partial \phi_0^2}+\frac{m^2a^4}{2}\phi_0^2+\sum_{k\neq 0}^{\infty}\paq{-\frac{1}{2}\frac{\partial^2}{\partial v_k^2}+\frac{\omega_k^2}{2}v_k^2}\right\}\Psi\pa{a,\phi_0,\{v_k\}}=0
\ee
\end{widetext}

with ${\mathcal H}_I\equiv {\mathcal H}_0$ and $\hat {\mathcal H}_k\equiv \frac{1}{2}\frac{\partial^2}{\partial v_k^2}+\frac{\omega_k^2}{2}v_k^2$ for $k\neq 0$. The above quantum equation will be the starting point in our approach. In Section 3 we shall return to the particular factor ordering employed in quantising the gravitational kinetic term.

\section{Decomposition of the Quantum System}
Finding the general solution of the  WdW equation (\ref{ WdW0}), even in the absence of perturbations, is a very complicated task due to the interaction between matter and gravity.\\
A set of approximate solutions can be found within a BO approach. However what we shall follow here is only in part a BO approach as we shall separate the homogeneous and inhomogeneous modes of the wave-function and implicitly consider the latter as a perturbation of the former. The BO approximation was originally introduced in order to simplify the Schr\"odinger equation of complex atoms and molecules \cite{BO} and has been applied successfully to the inflaton-gravity system. Within such a BO decomposition the semiclassical limit can be recovered straightforwardly. Moreover this approach treats the quantum-mechanical probability flux correctly to all orders, without violating unitarity \cite{K,BO-unit}.\\ 
It consists in factorising the wave-function of the universe $\Psi\pa{a,\phi_0,\{v_k\}}$ into the product
\be{BOdec}
\Psi\pa{a,\phi_0,\{v_k\}}=\Psi_0\pa{a,\phi_0}\prod_{k\neq 0}^\infty \chi_k\pa{a,v_k}
\ee
where $\Psi_0\pa{a,\phi_0}$ is the wave-function for the homogeneous inflaton-gravity sector and $\chi_k\pa{a,v_k}$ is that for each mode of MS perturbation field. Let us note that the wave-function of each mode $k$ depends parametrically on the conformal time $\eta$ and, in the semiclassical (zero order) limit, the evolution of the scale factor $a=a(\eta)$ fixes $\eta$ as a function of $a$.\\
On projecting out the WdW equation on $\chi\equiv \prod_{k\neq 0}^\infty \chi_k\pa{a,v_k}$ one is led to an equation for the homogeneous inflaton+gravity system wave-function of the form
\be{graveq}
\paq{\frac{1}{2\Mt^2}\frac{\partial^2}{\partial a^2}+\hat {\mathcal H}_I+\av{\hat {\mathcal H}_{MS}}}\tilde \Psi_0=
-\frac{1}{2\Mt^2}\langle\frac{\partial^2}{\partial a^2}\rangle\tilde \Psi_0
\ee
where 
\be{gder}
\Psi_0=e^{-i\int^{a} \mathcal{A} da'}\tilde\Psi_0,\; \chi=e^{i\int^{a} \mathcal{A} da'}\tilde\chi, \; \mathcal{A}=-i\langle\chi|\frac{\partial}{\partial a}|\chi\rangle
\ee
and $\av{\hat O}\equiv\langle\tilde \chi|\hat O|\tilde\chi\rangle$. Each mode is individually normalised by $\langle \chi_{k}|\chi_{k}\rangle=\int d v_k\chi_k^*\chi_k=1$. The compact notation in (\ref{graveq}) and (\ref{gder}) needs further explanation. Let us first note that 
\be{Hm}
\av{\hat {\mathcal H}_{MS}}=\sum_{k\neq 0}\langle \chi_k| \hat {\mathcal H}_k|\chi_k\rangle,
\ee
$\mathcal{A}=\sum_{k\neq 0} \mathcal{A}_k$ with $\mathcal{A}_k\equiv -i\langle\chi_k|\frac{\partial}{\partial a}|\chi_k\rangle$ and 
\be{tildechik}
\tilde\chi=\prod_{k\neq 0}\tilde \chi_k\;\,{\rm with}\;\,\chi_k=e^{i\int^{a} \mathcal{A}_k da'}\tilde\chi_k,
\ee
and finally
\be{deckgrav}
\langle\frac{\partial^2}{\partial a^2}\rangle=\sum_{k\neq 0} \langle \tilde\chi_k|\frac{\partial^2}{\partial a^2}|\tilde \chi_k\rangle.
\ee
On then neglecting the back-reaction of the quantum fluctuations on the homogeneous part, Eq. (\ref{graveq}) becomes
\be{graveqNOPER}
\paq{\frac{1}{2\Mt^2}\frac{\partial^2}{\partial a^2}+\hat {\mathcal H}_I}\tilde \Psi_0\simeq
0.
\ee
The equation for $\tilde\chi_k$ can be obtained by projecting out the WdW equation on $\prod_{j\neq k}^\infty \chi_j$ and is 
\ba{mateqk}
&&\tilde\Psi_0^*\tilde\Psi_0\paq{\hat {\mathcal H}_k-\langle \tilde \chi_k|\hat {\mathcal H}_{k}|\tilde \chi_k\rangle}\tilde \chi_k+\frac{1}{\Mt^2}\pa{\tilde\Psi_0^*\frac{\partial \tilde \Psi_0}{\partial a}}\nonumber\\
&&\times \frac{\partial \tilde \chi_k}{\partial a}=\frac{1}{2\Mt^2}\tilde\Psi_0^*\tilde\Psi_0\paq{\langle \tilde \chi_k|\frac{\partial^2}{\partial a^2}|\tilde \chi_k\rangle-\frac{\partial^2}{\partial a^2}}\tilde \chi_k.
\ea
This latter equation contains the wave-function of the homogeneous universe $\tilde \Psi_0(a,\phi_0)$ and the wave-function of a single $k$-mode. One needs to substitute the appropriate Hamiltonian in (\ref{mateqk}) in order to obtain the corresponding quantum evolution.\\
The right hand side (r.h.s.) of Eqs. (\ref{graveq},\ref{mateqk}) describes the non adiabatic transitions in our BO decomposition and are generally associated with the quantum gravitational effects (QGE).  
The homogeneous part of the wave function of the universe is further decomposed into the wave-function for the scale factor $\psi(a)$ and that for the homogeneous inflaton $\chi_0\pa{a,\phi_0}$ as
\be{psi0}
\tilde\Psi_0(a,\phi_0)\equiv \psi(a)\chi_0\pa{a,\phi_0}.
\ee
For $\chi_0(a,\phi_0)$, we shall consider a quantum state highly peaked on some value of $\phi_0$. This latter assumption is used both when considering a classical limit  and in the loop space formulation \cite{loop}. In  the current   approach we shall neglect the r.h.s. of  Eq. (15) and evaluate the effect of the above non classical homogeneous wave functions on the evolution of the MS variables.
If we consider a generic homogeneous matter wave function and again perform  a B.O. decomposition, as for the inhomogeneous part, together
with the semiclassical limit for gravitation we can introduce time.  One can then study the evolution of the MS variables and the effect of
the right hand side of Eq. (\ref{mateqk}) describing non adiabatic transitions in our BO decomposition, which are  generally associated with quantum gravitational effects (QGE). In particular these QGE affecting the evolution of the MS variables have been investigated in a series of paper \cite{BO-cosm,K,quantumloss} and have been shown to lead to $k$ a dependent deviation from the standard inflationary spectra generated during inflation.

\section{The Emergence of Time and Bounce}
The evolution of the MS fields depends on the wave-function of the homogeneous universe $\tilde \Psi_0(a,\phi_0)$. The homogeneous inflaton-gravity system, for the chaotic inflation case, has been studied in different papers \cite{TVV,BO-unit,FVV,FMVV}. In particular we are interested in the approach followed in \cite{TVV} wherein the system is solved in a inflationary regime with a highly peaked inflaton wave-function and correspondingly a highly oscillatory quantum state for gravity. Such an approach is very different from the more conventional method of considering the semiclassical limit for gravity and then studying the evolution of the wave-function for the inflaton field. In particular the latter approach naturally leads to the emergence of time which can be associated with the (quasi)-classical trajectory of the scale factor. Still in \cite{TVV} we showed that time evolution can be associated with the (quantum) probability flux even in the absence of a well defined classical trajectory (the homogeneous scalar field being quasi-classical, that is highly peaked, and time independent). Let us examine this approach in detail.\\
In what follows we consider the BO factorisation (\ref{psi0}) and the generic factor ordering \cite{order}
\be{ordering}
\partial_a^2\rightarrow a^{-i}\partial_a a^{-j}\partial_a a^{-k}\;\,{\rm with}\,\;i+j+k=0
\ee
leading from Eq. (\ref{graveqNOPER}) to the following homogeneous WdW equation :
\begin{widetext}
\begin{eqnarray}\label{PDE}
&\chi_0(a,\phi_0)\:\partial_{a}^2 \psi(a)+ 2\: \partial_{a}\chi_0(a,\phi_0)
\:\partial_{a}\psi(a)+\psi(a)\:\partial_{a}^2 \chi_0(a,\phi_0)+{}&\nonumber\\
&{}\frac{k(1+j+k)}{a^2}\psi(a)\chi_0(a,\phi_0)-\frac{j+2k}{a}\paq{\chi_0(a,\phi_0)\partial_a\psi(a)+\psi(a)\partial_a\chi_0(a,\phi_0)}+{}&\nonumber\\
&{}+2 \Mt^2 \psi(a)\hat{\mathcal H}_I \chi_0(a,\phi_0)=0.&
\end{eqnarray}
\end{widetext}
We now note that the Hamiltonian of the homogeneous inflaton, in the chaotic inflation framework, is that of an harmonic oscillator 
\be{Hinf}
\hat {\mathcal H}_{I}(a)\equiv\frac{\hat\pi_{\phi}^2}{2 a^2}+
\frac{m^2 a^4}{2}\hat\phi_0^2=m a \left( b^{\dagger} b+\frac{1}{2}\right)\, ,
\ee
where
\begin{eqnarray}
&&b=\sqrt{\frac{m a^3}{2}}\left(\hat\phi_0+\frac{i}{m a^3}
\hat\pi_{\phi}\right), \nonumber \\  
&&b^{\dag}=\sqrt{\frac{m a^3}{2}}\left(\hat\phi_0-
\frac{i}{m a^3}\hat\pi_{\phi}\right)
\label{bbstar}
\end{eqnarray}
and $m$ is the inflaton mass. 
\subsection{Solution for $a$ large}
Let us now consider the following ansatz for the inflaton state satisfying: 
\be{ansatz}
b|\chi_0\rangle=\alpha(a)|\chi_0\rangle
\ee
with $\alpha(a)=\sqrt{\frac{m \bar\phi^{2}a^3}{2}}$ and $\bar \phi$ free parameter. The ansatz (\ref{ansatz}) consists\footnote{some motivations can be given for such a choice as being associated with the random creation process of a large number of inflaton quanta \cite{TVV} around some mean (large) value.} of a coherent state for the inflaton corresponding to the following wave-function 
\begin{equation}\label{u}
\chi_0(a,\phi_0)=\left(\frac{m a^3}{\pi}\right)^{\frac{1}{4}}
\exp\left[-\frac{m a^3}{2}(\phi_0-\bar \phi)^2\right]\, ,
\end{equation}
which is a simple gaussian peaked around $\bar \phi$ with a width which decreases as 
$m a^3$ increases. The dependence on $a$ is chosen so as to obtain $\av{\mathcal H}_I\sim a^4 m^2\bar \phi^2/2$ ($a$ large) and a nearly constant energy density during inflation $\rho_I\sim m^2 \av{\hat \phi_0^2}/2$.\\
If we substitute the expression (\ref{u})
into the equation (\ref{PDE}) and calculate the contributions of
the different derivatives we obtain:
\begin{eqnarray}
&&\!\!\!\!\!\!\!\!\!\!\!\!\partial_{a} \chi_0=\frac{3}{4 a}\left[1-2 m a ^3
(\phi_0-\bar\phi)^2\right]\chi_0\label{est1}\, ,\\
&&\!\!\!\!\!\!\!\!\!\!\!\!\partial_{a}^2 \:\chi_0=-\frac{3}{16 a^2}\left[1-28 m a^3
(\phi_0-\bar\phi)^2\right.\nonumber \\
&&\left.+12 m^2 a^6(\phi_0-\bar\phi)^4
\right]\chi_0\label{est2}\, ,\\s
&&\!\!\!\!\!\!\!\!\!\!\!\!\frac{\partial^2}{\partial\phi_0^2}\chi_0=
-m a^3\left[1-m a^3
(\phi_0-\bar\phi)^2\right]\chi_0\label{est3}\, .
\end{eqnarray}
For $ma^3$ large 
\begin{eqnarray}
&&\max\left[ \pa{m a^3}^n|\phi_0-\bar\phi|^{2n} \chi_0(a,\phi_0)\right]\nonumber \\
&&\sim
\left(2n\right)^{n}
\exp\left[-n\right]\max \left[\chi_0(a,\phi_0)\right]\, ;
\label{dis}
\end{eqnarray}
and the contributions (\ref{est1}-\ref{est3}) are 
\begin{eqnarray}
&&\partial_{a} \chi_0(a,\phi_0)=O( a^{-1}) \; \chi_0(a,\phi_0)\, \label{u1eq},\\
&&\partial_{a}^2 \:\chi_0(a,\phi_0)=O(a^{-2}) \; \chi_0(a,\phi_0)\, \label{u2eq}, \\
&&\frac{\partial^2}{\partial\phi^2}\chi_0(a,\phi_0)=O(a^{3}) \; \chi_0(a,\phi_0) \;
\end{eqnarray}
and are thus subleading for $a$ large.
On just retaining the leading contributions in (\ref{PDE}) one finally has:
\begin{equation}\label{leadPDE}
\left[\partial_{a}^2 \psi(a)+m^2 \Mt^2 \phi_0^2
\:a^4 \psi(a)\right]\: \chi_0(a,\phi_0)=0
\end{equation}
where $\chi_0(a,\phi_0)$ has support in a tiny region around $\bar\phi$, due to the
large values of $a$. Let us note that the result obtained in the inflationary limit is independent of the ordering as previously pointed out \cite{FVV}. 
Finally one may rewrite (\ref{leadPDE}) as:
\begin{equation}\label{qairy}
\partial_{a}^2 \psi(a)+m^2 \Mt^2 \bar\phi^2\: a^4 \psi(a)=0.
\end{equation}
A general solution in terms of the Bessel functions $N_{\nu}$ and $J_{\nu}$
of Eq. (\ref{qairy}) is 
\begin{equation}
\psi(a) = \sqrt{a}\left[C_1N_{\frac16}\left( \frac{m \Mt \bar\phi}{3}a^{3}\right)+
C_2J_{\frac16}\left( \frac{m \Mt \bar\phi}{3}a^{3}\right)\right]
\label{Bessel}
\end{equation}
and for $a\rightarrow \infty$ with $D_i = C_i\sqrt{\frac{12}{\pi m \Mt \bar\phi}}$: 
\begin{equation}\label{genAIRYa}
\psi(a)\sim a^{-1}\left[D_{1}\sin \frac{m \Mt \bar\phi}{3}
a^{3}+D_{2}\cos \frac{m \Mt \bar\phi}{3}a^{3}\right],
\end{equation}
 where $D_1$ and $D_2$ are complex numbers.
The oscillatory behavior is encoded in $\psi$, even if at an
approximate level (the solution is not exact).
Let us note that the $D_i$ ($C_i$) in  can be determined by
the initial conditions. For example, $C_1=-i$ and $C_2=1$ corresponds to
the Vilenkin initial wave-function of the universe, while if
$C_1=1$ and $C_2=0$ one has the Hartle-Hawking choice \cite{UNI}. \\
The (approximate) solution of the homogeneous equation can then be found as the product of a gaussian function peaked on a constant $\bar \phi$ and a highly oscillating function of the scale factor. In such a case the wave-function for $a$ is non classical but time can be still introduced. 

Each mode of the MS fields is described by a wave-function $\chi_k$ which satisfies Eq. (\ref{mateqk}). On neglecting the r.h.s., Eq. (\ref{mateqk}) has the form of a Schr\"odinger-like equation where the time derivative is replaced by
\begin{eqnarray}
&&\frac{1}{\Mt^2}\pa{\tilde\Psi_0^{-1}\frac{\partial \tilde \Psi_0}{\partial a}}\partial_a\chi_k(a,v_k)\nonumber \\
&&=\frac{1}{\Mt^2}\pa{\psi^{-1}\frac{\partial \psi}{\partial a}+\chi_0^{-1}\frac{\partial \chi_0}{\partial a}}\partial_a\chi_k(a,v_k)
\label{timeder}
\end{eqnarray}
and the contribution $\chi_0^{-1}\frac{\partial \chi_0}{\partial a}$ is negligible.\\
In the semiclassical limit (WKB) with $\psi(a)=\pa{a'_{cl}}^{-1/2}\exp\paq{{-i\int_a a'_{cl}da}}$ the probability density is inverse proportional to the velocity calculated on the classical trajectory and (\ref{timeder}) has support on the classical trajectory and no coarse graining is needed in order to connect the solution to the classical regime. In such a case one has
\be{wkbtime}
\frac{\partial_a \psi}{ \psi}\sim -i a'_{cl}\equiv-i h_{cl} a^2
\ee
which properly defines a ``classical'' time $\eta$ with $a'_{cl}\partial_a=\partial_\eta$ and the Hubble parameter $h_{cl}$. For the general solution (\ref{genAIRYa}) time emerges after coarse graining over one period $\Delta a\sim \frac{2 \pi}{m \Mt \bar\phi a^{2}}$ and one finds
\begin{equation}\label{CG}
\frac{\partial_{a}\psi}{\psi}\sim\pm i\,
m \Mt \bar\phi\, a^{2}
\end{equation}
or zero depending on the values of the integration constants $D_1$ and $D_2$. Let us note that on coarse graining the full matter equation over one (short) period of oscillation of the gravitational wave-function one neglects small contributions which arise due to the dependence of the matter wave-function on $a$. Coarse graining is not even necessary if $D_1/D_2=\pm i$ as oscillation disappears and one obtains (\ref{CG}) without averaging the matter equation.\\
The presence of an effective time evolution for matter arises from a mechanism similar to one already observed in the analysis of the classical limit of quantum systems, such as the hydrogen atom [15], in the sense that the quantum probability as a function of $a$ is similar to the measure of temporal density in a classical orbit. This fact has been studied for the stationary quantum eigenstates of the hydrogen atom (with two particular fixed values of the angular momenta and large principal quantum number n) some of which presents a radial highly oscillatory behaviour. On course graining (in particular on applying the Riemann-Lebesgue Lemma) one is able to recover the classical trajectory related to the given angular momenta. Indeed the classical trajectory is related to a classical spatial probability distribution of a particle in terms of the inverse of its speed (the fraction of time spent in a spatial interval is a measure of the probability density).\\
We note that different choices for operator ordering in the WdW equation may lead to effects at the beginning of the inflationary phase and could modify correspondingly quantum gravitational contributions to the power spectrum \cite{K}, we hope to return to this. 
\subsection{Solution for $a$ small}
We have found a solution to our homogeneous system for $a$ large. Let us nonetheless use it also for $a$ small. If we add the additional hypothesis that the exponent of our gaussian (\ref{u}) tends to zero for $a$ small, implying that $\phi_0$ is never too large (no random creation of too large a number of inflaton quanta \cite{TVV}) one just obtains, on retaining the leading terms,
\begin{eqnarray}
&&\partial_a^2\psi+\frac{1}{a}\pa{\frac{3}{2}+i-k}\partial_a\psi\nonumber \\
&&-\frac{1}{a^2}\paq{\frac{3}{16}+k\pa{i-1}+\frac{3}{4}\pa{i-k}}\psi=0.
\label{Bgianni}
\end{eqnarray}
At this point in analogy with Eq. (\ref{wkbtime}) it is convenient to introduce, for $a$ small,
\be{timeder2}
\frac{\partial_a\psi}{\psi}=\frac{c}{a}=a^2\pa{\frac{c}{a^3}}
\ee
where the term $c/a^3$ is, for $a$ large, replaced by $\pm i m\Mt \bar\phi$ which is related to the value of the Hubble constant (see Eq. (\ref{CG}) and Ref. \cite{TVV}) and determines it. From Eq. (\ref{Bgianni}) one obtains
\begin{eqnarray}
&&a\partial_a c+c^2+c\pa{\frac{1}{2}+i-k}\nonumber \\
&&-\paq{\frac{3}{16}+k\pa{i-1}-\frac{3}{4}\pa{i-k}}=0.
\label{Bgianni2}
\end{eqnarray}
and one has a solution for $c$ constant and 
\be{Bgianni3}
c=\frac{1}{2}\paq{-\pa{\frac{1}{2}+i-k}\pm \sqrt{3-4k+\pa{i+k}^2}}
\ee
which of course is real for the simplest case $i=j=k=0$, also for the Laplace-Beltrami choice $i=1$, $j=-1$, $k=0$ and the Vilenkin choice $i=-1$, $j=1$, $k=0$ \cite{order}. The result suggests that, as $a$ becomes small, the Hubble parameter eventually becomes imaginary, implying that the universe will stop contracting and is suggestive of a bounce. Indeed in Eq. (\ref{CG}) one has two possible signs for the time: one associated with a contracting and the other with and an expanding universe and presumably the two ``meet'' for $a$ small when the Hubble parameter (or time) becomes imaginary. Admittedly the above approach to a possible bounce is very heuristic, however its possible occurrence, with the necessary presence of matter, should not surprise us. Indeed in the usual quantum mechanics for potentials which are dominated by the centrifugal one near the origin, there is no collapse and it is the latter that determines the behaviour of the wave-function there (see for example \cite{Landau}). \\
Let us improve on the above approach by now making the following again highly peaked ansatz for the matter wave-function
\be{ansB}
u(a,\phi_0)=\pa{\frac{2\beta}{\Mt^2\pi}}^{1/4}\exp\paq{-\frac{\beta}{\Mt^2}\pa{\phi_0-\phi_a}^2},
\ee
where $\phi_a$ is a function of $a$ to be calculated. The full matter-gravity wave-function is $\Psi(a,\phi_0)\equiv\psi(a)u(a,\phi_0)$ and the WdW can be written as the sum of three contributions which must be zero. The contribution proportional to $\phi_0^2$ is given by
\be{phi2c}
m^2\Mt^2a^4-\frac{4\beta^2}{\Mt^2a^2}+4\frac{\beta^2}{\Mt^4}\phi_a'^2=0,
\ee
where the prime denotes the derivative w.r.t. $a$. For $\beta\gg m\Mt^2 a^3$ the first contribution can be ignored and one finds an equation for $\phi_a$ with the following solution
\be{phias}
\phi_a=\bar \phi+\Mt \ln a/a_0.
\ee
The second contribution is proportional to $\phi_0$ and must be set independently to zero
\be{phi1c}
\frac{2 \beta}{\Mt a^2}\pa{2a\frac{\partial_a\psi}{\psi}-j-2k-1}=0
\ee
leading to
\be{gravsmall}
\psi=B a^{\frac{1}{2}\pa{1+j+2k}}. 
\ee
Finally one is left with the third contribution
\begin{eqnarray}
&&\frac{1}{\beta a^2}\pa{k+jk+k^2}+\frac{2\pa{1+j+2k} \phi_a}{a^2\Mt}\nonumber \\
&&-\frac{\psi'}{a\psi}\frac{\pa{j+2k}\Mt+4\beta \phi_a}{\beta\Mt}+\frac{1}{\beta}\frac{\psi''}{\psi}=0,
\label{phic}
\end{eqnarray}
where we divided the equation by $\beta$. Let us note that for $\beta$ large  and in particular $\phi_a/\Mt\gg 1/\beta$ the leading contributions in (\ref{phic}) are those proportional to $\beta^0$. Given the constraints (\ref{phias}), (\ref{gravsmall}) previously found, Eq. (\ref{phic}) finally becomes
\be{phics}
-\frac{1}{4a^2}\pa{k-1}^2=0
\ee 
and is satisfied for each ordering with $k=1$. However, if we leave the ordering unspecified and for $\beta\gg \Mt/\phi_a$, we note that Eq. (\ref{phic}) is still satisfied to the leading order.
We conclude that the wave-function
\begin{widetext}
\be{psisol2}
\Psi=B a^{\frac{1}{2}\pa{1+j+2k}}\pa{\frac{2\beta}{\Mt^2\pi}}^{1/4}\exp\paq{-\frac{\beta}{\Mt^2}\pa{\phi-\bar \phi-\Mt \ln a/a_0}^2}
\ee
\end{widetext}
is an approximate solution to the WdW equation in the regime with $\beta\gg 1\gg m\Mt^2 a^3$ and thus for $a$ small. \\
One can finally merge the solution obtained for $a$ small starting from (\ref{ansB}) and the one already proposed for $a$ large (\ref{u}).\\
Merging leads to an approximate solution which satisfies the WdW homogeneous equation for $a$ large and $a$ small starting from the following general ansatz for the matter wave-function
\begin{eqnarray}
&&u(a,\phi)=\pa{\frac{2\alpha(a)}{\pi}}^{1/4}\exp\paq{-\alpha(a)\pa{\phi-\phi_a}^2}\nonumber \\
&&\equiv \pa{\frac{2\alpha(a)}{\pi}}^{1/4}\exp\paq{-\alpha(a)\Delta\phi^2},
\label{ansBM}
\end{eqnarray}
where $\alpha=\frac{\beta}{\Mt^2}+\frac{ma^3}{2}$ and $\phi_a=\bar \phi+\ln a/\bar a$. For $a$ small one recovers the solution (\ref{psisol2}) with a non-evolving gravity wave-function (\ref{gravsmall}) and for $a$ large one finds (\ref{genAIRYa}) with $\bar \phi\rightarrow \phi_a$. We already discussed the emergence of time in the large $a$ regime. In such a context the Hubble parameter is defined contextually (see (\ref{wkbtime}) and (\ref{CG})) with $\partial_a\psi/\psi$ being an imaginary function of $a$. For $a$ small the same procedure would lead to a real ratio $\partial_a\psi/\psi$ or, correspondingly, to an imaginary Hubble parameter. Classically a complex Hubble parameter results as the solution of some exotic Friedmann equation describing a bouncing universe, its imaginary part being different from zero in the region which is classically not accessible. Analogously, at the quantum level, the wave-function satisfying the time-independent Schr\"odinger equation in the classically forbidden region is real and decreasing while it is oscillatory in the region classically accessible. We then argue that (\ref{ansB}) describes a bouncing universe and the different signs in (\ref{CG}) are presumably associated with its contracting (plus) and expanding (minus) phases. Unfortunately, within our approach, we are not able to calculate the turning point with precision. However we can estimate $a_B$, namely the value of the scale factor at the bounce, as the point where the approximations used to obtain the small $a$ solution break down. This happens for  $\beta\sim m\Mt^2 a^3$ and $\beta\sim \Mt/\phi_a$. Given the constraint coming from the observed amplitude of temperature fluctuations of the CMB $h_*/\Mt\sim 10^{-5}$ where $h_*\sim m\phi_*/\Mt$ is the classical Friedmann equation during inflation and assuming $\phi_a\sim \phi_*$ then we find
\be{a0bounce}
a_B\M\sim 2\cdot 10
\ee
i.e. $a_B$ is of the order of few times the Planck length.\\
In any case we argue that the QGE from such a bouncing scenario can be estimated by considering arbitrary superpositions of the wave-functions for gravity (different choices of $D_{1,2}$ in (\ref{genAIRYa})). The rest of this paper concentrates in such estimates and their possible effects on the inflationary spectra. \\
Let us comment on the significance of the presence of a ``bounce''. Its  presence can be associated  with a time-symmetric expansion followed by a contraction or vice-versa. It is only at the position of the bounce that expanding and contracting universes interfere (see footnote 8 in \cite{BO-cosm0} and \cite{cash}). Thus the Hartle-Hawking  no boundary initial condition, leading  to a real wave function of the universe, is associated with a time symmetric evolution whereas the Vilenkin choice corresponds to the  choice of one time branch with respect to the other or if we wish a tunnelling from nothing to de Sitter. Thus on choosing a ``mixed'' initial  condition we are introducing a ``small'' measure of time-reversal invariance violation. Let us also observe that a bounce (at Planck densities) has also been observed in the context of loop space 
quantum gravity \cite{loop}, which, as we have previously noted \cite{BO-unit} bears some resemblance to our present approach.
\section{Quantum Gravitational Effects}
In the WdW framework the homogeneous inflaton-gravity system plays a central role in determining the dynamics of the inflationary epoch and is also responsible for the emergence of time which parametrises the evolution of inhomogeneities (structures). The quantum origins of time may then have observable effects on the primordial spectra which one can evaluate. At the end of the previous section we mentioned the possibility of non trivial superpositions in the gravitational wave function arising because of the bounce or a small measure of time reversal invariance violation.\\
The equation governing the evolution of each $k$-mode of the MS field is (\ref{mateqk}) and in a series of papers \cite{K} we already studied the quantum gravitational effects originated by the r.h.s. of this equation. In this article these r.h.s. contributions are neglected as a first approximation and we limit ourselves to the analysis of the effects arising from different choices of the gravitational wave-function.\\ 
Our starting point is then
\begin{widetext}
\be{matk}
\paq{\hat {\mathcal H}_k-\langle \tilde \chi_k|\hat {\mathcal H}_{k}|\tilde \chi_k\rangle}\tilde \chi_k=-\frac{1}{\Mt^2}\pa{\frac{\partial_a\psi}{\psi}+\frac{\partial_a \chi_0}{\chi_0}} \frac{\partial \tilde \chi_k}{\partial a}\sim -\frac{1}{\Mt^2}\frac{\partial_a\psi}{\psi} \frac{\partial \tilde \chi_k}{\partial a}
\ee
where 
\be{chineg}
\frac{\partial_a \chi_0}{\chi_0}=\mathcal{O}(a^{-1})
\ee
is negligible for $a$ large and 
\be{dpsiopsi}
\frac{\partial_a\psi}{\psi}=3\Omega_0^3a^2\frac{D_1\cos\pa{\Omega_0a}^3-D_2\sin\pa{\Omega_0a}^3}{D_1\sin\pa{\Omega_0a}^3+D_2\cos\pa{\Omega_0a}^3}+\mathcal{O}(a^{-1})
\ee
\end{widetext}
with $\Omega_0\equiv \pa{m\Mt\bar\phi/3}^{1/3}$ and $D_{1,2}$ are complex numbers. Let us note that if $D_1/D_2=\pm i$ and neglect the contributions $\mathcal{O}(a^{-1})$ one has 
\be{dpsiopsi2}
\frac{\partial_a\psi}{\psi}=\pm i m\Mt\bar\phi \,a^2,
\ee
even without coarse graining and the minus sign corresponds to a parametrisation which associates an increasing time to the expansion of the universe. Let us now consider the case $D_1/D_2=-i+\epsilon$ which corresponds to a choice of initial conditions being a superposition of Vilenkin and Hartle-Hawking ones thus introducing possible effects of interference between contracting and expanding universes. In such a case, if $\epsilon\ll 1$ one has a residual phase 
\be{dpsiopsi3}
\frac{\partial_a\psi}{\psi}=- i m\Mt\bar\phi \,a^2\paq{1+\epsilon \exp\pa{\frac{2}{3}im\Mt\bar\phi\,a^3}},
\ee
which encodes the effects arising from the quantum origin of time. Further such effects can be treated perturbatively. The leading order contribution in (\ref{dpsiopsi3}) is responsible for the emergence of time through
\be{time0}
i\frac{m\bar \phi a^2}{\Mt}\partial_a\equiv i\frac{\rd}{\rd\eta}
\ee
and we treat the additional (varying) phase as a perturbation. This latter relation also determines the ``classical trajectory'' for the scale factor $a(\eta)$. Finally Eq. (\ref{matk}) takes the following form:
\be{mateqks}
i\partial_\eta\chi_s-\hat{\mathcal{H}}_k\chi_s=- \epsilon \exp\pa{\frac{2}{3} i m\Mt\bar\phi\,a^3}\pa{i\partial_\eta-\av{\hat{\mathcal H}}_s}\chi_s,
\ee
where 
\be{defchis}
\chi_s\equiv\chi_k\exp\pa{i\int_\eta\av{\hat{\mathcal H}_k(\eta')}d\eta'},\; \av{\hat{\mathcal H}}_s\equiv\langle \chi_s|\hat{\mathcal H}_k|\chi_s\rangle
\ee
and the ``classical'' dependence of $a$ on $\eta$, defined implicitly by (\ref{time0}), has been used. The r.h.s. of this latter equation encodes the QGE originated by a superposition of the solutions of the Schr\"odinger-like equation for homogeneous gravity (\ref{qairy}). Such effects were not studied in our previous approach \cite{K} and were neglected in \cite{TVV} where we made the assumption that, for $a$ large the r.h.s. simply averages to zero. Such an assumption, however, must be verified a posteriori since for certain values of the parameters or at some time during the evolution of the primordial fluctuation these effects may have relevant consequences on the shape of the primordial spectra originating from inflation. Let us note that $\epsilon$ is simply related to the integration constants $D_{1,2}$ and can be negligible. In such a case these QGE are unobservable.\\
The state $|\chi_s\rangle$ satisfying Eq. (\ref{mateqks}) can be evaluated by standard perturbation theory: let $|0\rangle$ be the Bunch-Davies (BD) vacuum \cite{BD} which satisfies the unperturbed equation ($\epsilon\rightarrow 0$), then the perturbed vacuum can be written as the superposition of the full set of solutions of the unperturbed equation $|n\rangle$:
\be{pertBD}
|\chi_s\rangle=|0\rangle+\epsilon \sum_{n\neq 0} c_n(\eta)|n\rangle.
\ee
On inserting this expression in (\ref{mateqks}) and just keeping the first order in $\epsilon$ one finds the following differential equation for the amplitudes $c_n(\eta)$
\be{pertBDcn}
i\partial_\eta c_n=\exp\pa{\frac{2}{3} i m\Mt\bar\phi\,a^3}\langle n| \pa{\hat{\mathcal H}_k-\langle 0|\hat{\mathcal H}_k|0\rangle}|0\rangle.
\ee
\subsection{The Unperturbed Case}
Here we briefly review how the unperturbed equation obtained in the $\epsilon\rightarrow 0$ limit of (\ref{mateqks}) can be formally solved in terms of a time-dependent quantity called Pinney variable. Such a treatment \cite{FVV,Erm-Pin} is valid in general for Schr\"odinger-like equations with a time-dependent Hamitonian $\hat{\mathcal H}$ and consists in finding an ``invariant operator'' satisfying the equation
\be{inveq}
i \frac{d}{d\eta}\hat I+\paq{\hat I,\hat{\mathcal H}}=0.
\ee
The properly rephased eigenstates of the invariant are solutions of the Schr\"odinger equation. In particular if one is able to find two linear invariants $\hat I$ and $\hat I^\dagger$ satisfying the usual algebra of the creation-annihilation operators, then the complete set of the solutions can be built starting from the invariant vacuum state defined as $\hat I|0\rangle=0$.
The complete basis of solutions can then be generated by $\hat I^\dagger$ and its elements labelled by integer numbers.\\
In our case the invariant has the following form
\be{InvP}
\hat I=\frac{e^{i\Theta}}{\sqrt{2}}\paq{\pa{\frac{1}{\rho}-i\rho'}\hat v+i\rho\hat \pi}
\ee
where $\hat v$ is the MS variable, $\hat \pi$ is its conjugate momentum and $\Theta=\int^{\eta} \frac{d\eta'}{\rho^{2}}$. The Pinney variable $\rho$ satisfies the following non-linear differential equation (the so-called Ermakov--Pinney (EP) equation \cite{Erm-Pin})
\be{pin}
\rho''+\omega^{2}\rho=\frac{1}{\rho^{3}}
\ee
with $\omega^2=k^2-z''/z$ for the scalar MS variable and $\omega^2=k^2-a''/a$ for the tensor case. In the coordinate representation, the properly normalised BD vacuum, expressed in terms of the Pinney variable, is
\be{vacBD}
\langle v|0\rangle_s=\frac{1}{\pa{\pi\rho^2}^{1/4}}\exp\paq{-\frac{i}{2}\int^\eta \frac{d\eta'}{\rho^2}-\frac{v^2}{2}\pa{\frac{1}{\rho^2}-i\frac{\rho'}{\rho}}}.
\ee
Let us finally note that the two-point function $p\equiv\langle 0|\hat v^2|0\rangle$ is given by
\be{pun}
p(\eta)=\frac{\rho^2}{2}
\ee
in terms of the Pinney variable. 
\subsection{The Perturbed Case}
In order to evaluate the matrix element $\langle n| \hat{\mathcal H}_k|0\rangle$ in Eq. (\ref{pertBDcn}) one may conveniently express $ \hat{\mathcal H}_k$ in terms of the invariants $\hat I$ and $\hat I^\dagger$:
\begin{widetext}
\ba{Hinv}
\hat{\mathcal H}_k=\frac{1}{4}\left\{\paq{\pa{\rho'+\frac{i}{\rho}}^2+\omega^2\rho^2}\re^{2i\Theta}\pa{\hat I^\dagger}^2+\paq{\pa{\rho'-\frac{i}{\rho}}^2+\omega^2\rho^2}\re^{-2i\Theta}\hat I^2+\paq{\rho'^2+\frac{1}{\rho^2}+\omega^2\rho^2}\pa{1+2\hat N}
\right\},
\ea
where $\hat N\equiv \hat I^\dagger\hat I$. This last expression can be then rewritten in terms of $\langle 0|\hat{\mathcal H}_k|0\rangle\equiv E_0$ in more compact form as
\ba{Hinv2}
\hat{\mathcal H}_k&=\left\{\paq{E_0-\frac{1}{2\rho^2}+i\frac{\rho'}{2\rho}}\re^{2i\Theta}\pa{\hat I^\dagger}^2+E_0\pa{\hat N+\frac{1}{2}}+{\rm h.c.}
\right\}.
\ea
Eq. (\ref{pertBDcn}) then takes the form
\be{pertBDcn2}
i\partial_\eta c_n=\delta_{n,2}\sqrt{2}\exp\pa{\frac{2}{3} i m\Mt\bar\phi\,a^3}\paq{E_0-\frac{1}{2\rho^2}+i\frac{\rho'}{2\rho}}\re^{2i\Theta}
\ee
where $\delta_{n,2}$ is the Kronecker delta. To the first order in $\epsilon$ one is then left with a single contribution
\be{c2}
c_2=\int_{\eta_0}^{\eta}\sqrt{2}\paq{E_0-\frac{1}{2\rho^2}+i\frac{\rho'}{2\rho}}\re^{2i\paq{\Theta+\pa{\Omega_0 a}^3}}d\eta'
\ee 
or
\be{c2a}
c_2=\frac{\sqrt{2}\Mt^2}{3\Omega_0^3}\int_{a_0}^{a}\paq{E_0-\frac{1}{2\rho^2}+3i\frac{\Omega_0^3 \bar a^2}{\Mt^2}\frac{\partial_{\bar a}\rho}{2\rho}}\re^{2i\paq{\Theta+\pa{\Omega_0 \bar a}^3}}{\bar a}^{-2}d\bar a.
\ee 
\end{widetext}
where $\eta_0$ or, correspondingly $a_0$, denotes the ``beginning'' of inflationary phase or, in terms of the solution of the homogenous WdW equation (\ref{ansBM}), that of the ``$a$ large regime''.
The function which must be integrated in order to calculate $c_2$ has an oscillatory behaviour. If the frequency of oscillation is high one expects that the integral averages to zero (Riemann-Lebesgue). Consequently, in the interval $[a_0,a]$, the transition amplitude $c_2$ receives the major contributions when the frequency of oscillation has the minimum value.\\
Given, at least formally, the expression for $|\chi_s\rangle$ one finally must calculate the power of primordial spectra which is proportional to the quantity $\langle\chi_s|\hat v^2|\chi_s\rangle$. If we express $\hat v^2$ in terms of $\hat I$, $\hat I^\dagger$ we finally obtain
\be{v2p}
\av{\hat v^2}=\frac{1}{2}\paq{\re^{-2i\Theta}\av{\hat I^2}+\av{\hat N+\frac{1}{2}}+{\rm c.c.}}
\ee
where the averages are calculated in terms of the perturbed $|\chi_s\rangle$. The result as a function of $c_2$ is
\be{v2p2}
\av{\hat v^2}=\frac{\rho^2}{2}\paq{1+\epsilon\sqrt{2}\pa{c_2{\rm e}^{-2i\Theta}+c_2^* {\rm e}^{2i\Theta}}}\equiv\frac{\rho^2}{2}\paq{1+\Delta_k}.
\ee

\begin{figure}[t!]
\centering
\epsfig{file=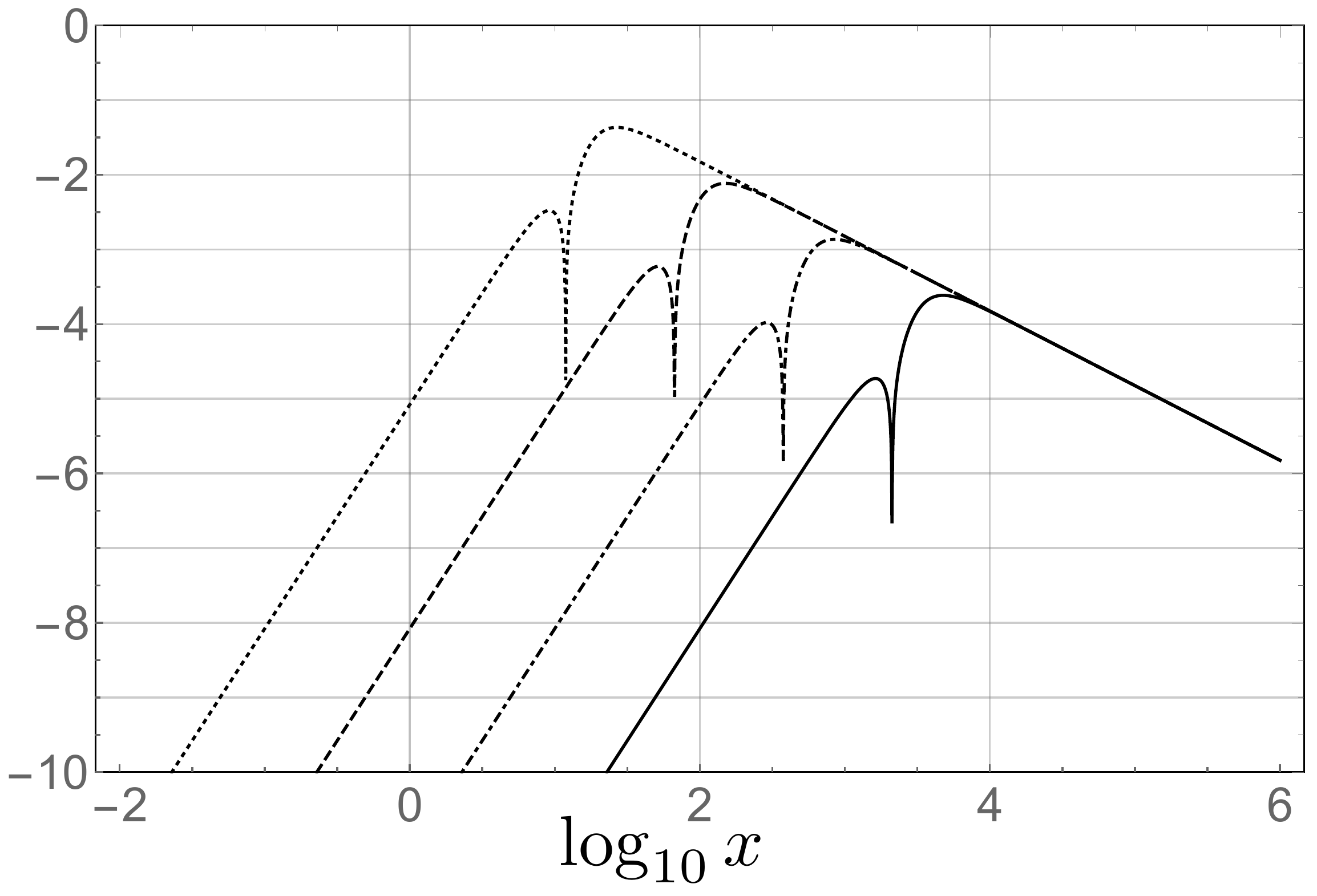, width=6 cm}
\epsfig{file=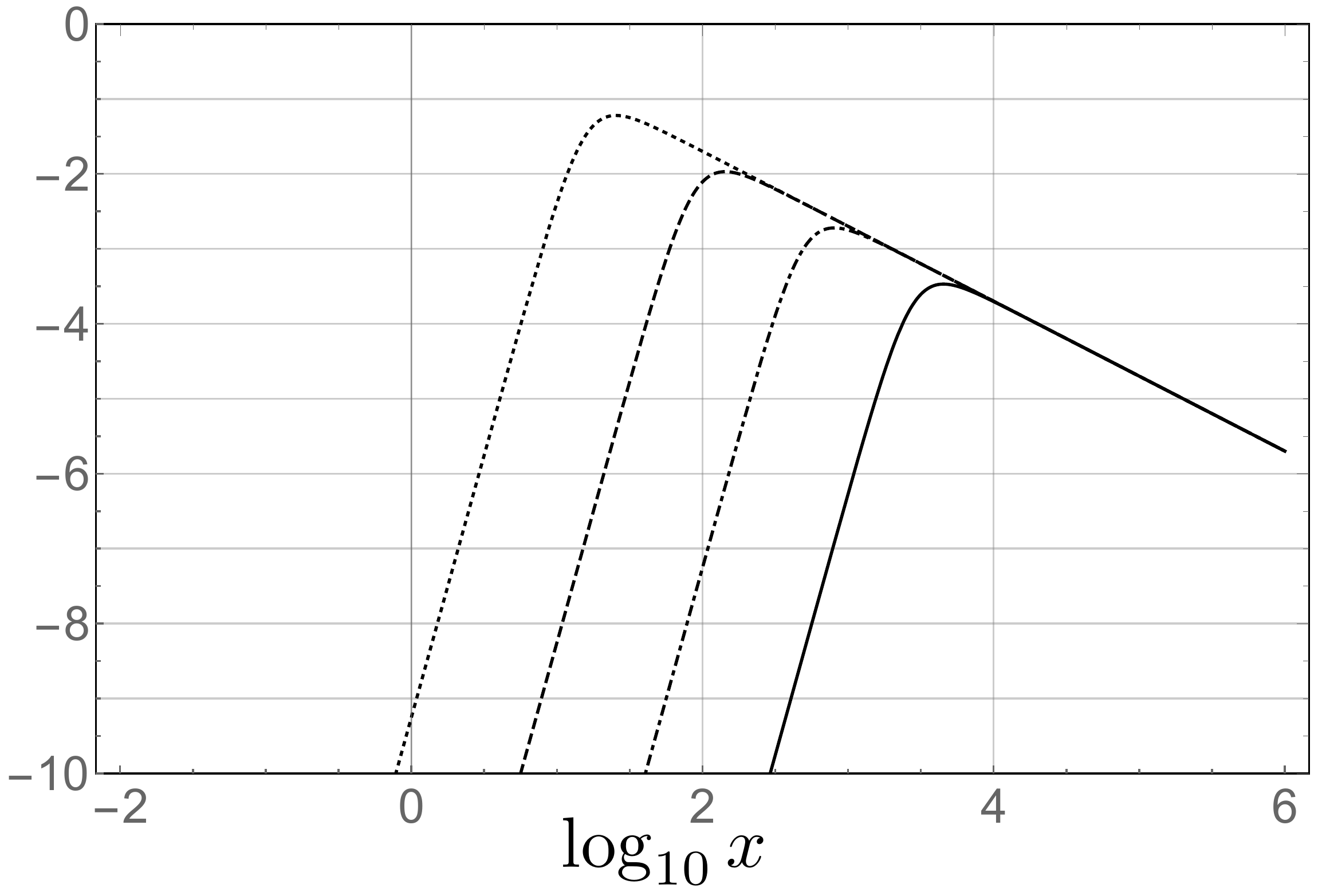, width=6 cm}
\caption{\it In the figure above we plotted the expression (\ref{R}) for $n=3$ (on the left) and $n=4$ (on the right). The dotted, dashed, dot-dashed and solid lines represent the four different choices $\frac{\M^2}{h^2}=10^4,10^7,10^{10},10^{13}$ respectively.}
\label{intbyparts}
\end{figure}
\section{De Sitter Case}
In this last section we shall estimate the QGC to the inflationary spectra (\ref{v2p2}) in the de Sitter (dS) limit. Such a limit represents a good approximation to the inflationary phase when, at least at the semiclassical level, the Hubble parameter is slowly varying (in the framework of slow-roll inflation). Moreover the spectra of the scalar and the tensor fluctuations are the same and obey the MS equation
\be{MSeq}
v''+\pa{k^2-\frac{2}{\eta^2}}v=0.
\ee
In the dS limit both the MS and the EP equations can be solved exactly. In particular on assuming a Bunch-Davies boundary condition for the vacuum state one has
\be{rhods}
\rho^2=\frac{1+k^2\eta^2}{k^3\eta^2}. 
\ee
where the conformal time can be easily rewritten in terms the scale factor as $\eta=\pa{a h}^{-1}$ where $h=a'/a^2={\rm const}$. With the above results one can then calculate the dynamical phase
\be{THds}
\Theta=\int_{\eta_0}^\eta\frac{d\eta'}{\rho^2}=k\pa{\eta-\eta_0}-\arctan\pa{k\eta}+\arctan\pa{k\eta_0}
\ee
and
\be{As}
A^*\equiv E_0-\frac{1}{2\rho^2}+i\frac{\rho'}{2\rho}=\frac{\pa{1-2i k\eta}\pa{k\eta-i}}{k^3\eta^4\pa{k\eta+i}}
\ee
with
\be{Ae2ith}
A^*\re^{2i\Theta}=\frac{k\eta_0-i}{k\eta_0+i}\frac{1-2ik\eta}{k^3\eta^4}\re^{2ik\pa{\eta-\eta_0}}\equiv \frac{1-2ik\eta}{k^3\eta^4}\re^{2i\pa{k\eta+\alpha_0}}.
\ee
On now combining the contributions to the integrand we find
\be{c2ds}
c_2=\sqrt{2}\re^{2i\alpha_0}\int_{\eta_0}^{\eta}\frac{1-2ik\bar\eta}{k^3\bar \eta^4}\re^{2ik\bar\eta-4i \frac{\M^2}{h^2}\bar \eta^{-3}}d\bar \eta
\ee 
where we used the classical Friedmann equation $h^2=m^2\bar \phi^2/\Mt^2$ (neglecting the kinetic term for the inflaton) in the definition of $\Omega_0$. Let us note that the phase must also contain a positive factor arising from the volume of the flat 3-space (see the discussion after Eq. (\ref{act})) which in principle is unknown a priori. We reintroduce this factor in the following expression for the phase of (\ref{c2ds}) as $\bar k^3\equiv L^{-3}$
\be{alpha}
\alpha(\eta)=2k\eta-4 \frac{\M^2}{h^2}\pa{\bar k\eta}^{-3}.
\ee
The total phase (\ref{alpha}) is the sum of a negative dynamical contribution (also present in the semiclassical approach) and a positive quantum gravitational contribution generated by the superposition of two different states of the gravitational wave function.
In what follows we assume $\bar k=k_*$ where $k_*$ is the pivot scale, and the modes we are interested in (those we observe today in the anisotropies of the CMBR) are quite close to it ($k\sim k_*$). If one now redefines the integration variable $-k\eta=x$ the integral (\ref{c2ds}) takes the form
\begin{widetext}
\be{c2dsx}
c_2=-\sqrt{2}\re^{2i\alpha_0}\pa{\int_{x_0}^{x}\frac{\re^{-2i\bar x+4i \frac{\M^2}{h^2}\frac{k^3}{\bar k^3}\bar x^{-3}}}{\bar x^4}d\bar x +2i\int_{x_0}^{x}\frac{\re^{-2i\bar x+4i \frac{\M^2}{h^2}\frac{k^3}{\bar k^3}\bar x^{-3}}}{\bar x^3}d\bar x}
\ee 
where $\M/h\sim 10^5$ (this is a conservative assumption valid in most single field inflationary models based on the slow-roll paradigm) and $\frac{\bar k}{k}\sim 1$. Let $x_0$ be the moment when perturbation modes are well inside the horizon and assume that they are in the BD vacuum state with the dynamical phase much larger then the quantum gravitational one. In such a case $x_0\gg 1$. When $x$ decreases and the modes get closer to the horizon exit, at some time, given our hypothesis, the quantum gravitational contribution in the total phase begins to dominate. 
The integral in (\ref{c2dsx}) cannot be computed exactly but it can be well approximated by an ``integration by parts'' method (see \cite{bender}) which, to the leading order gives
\be{intbypart}
\int_{x_0}^{x}\frac{\re^{-2i\bar x+4i \frac{\M^2}{h^2}\frac{k^3}{\bar k^3}\bar x^{-3}}}{\bar x^n}d\bar x\simeq\left.\frac{\re^{-2i \bar x+4i \frac{\M^2}{h^2}\frac{k^3}{\bar k^3}\frac{1}{\bar x^{3}}+i\frac{\pi}{2}}
}{2\,\bar x^n\pa{1+\frac{6}{\bar x^4}\frac{\M^2}{h^2}\frac{k^3}{\bar k^3}}}\right|_{x_0}^{x}
\ee
where contributions of order $\pa{1+\frac{6}{x^4}\frac{\M^2}{h^2}\frac{k^3}{\bar k^3}}^{-2}$ have been neglected. 
In particular the remaining integral 
\be{rest}
\int_{x_0}^x \frac{d}{dx}\paq{2\,\bar x^n\pa{1+\frac{6}{\bar x^4}\frac{\M^2}{h^2}\frac{k^3}{\bar k^3}}}^{-1}\re^{-2i \bar x+4i \frac{\M^2}{h^2}\frac{k^3}{\bar k^3}\frac{1}{\bar x^{3}}+i\frac{\pi}{2}}d\bar x
\ee
\end{widetext}
has been neglected.
In order to verify this latter approximation in the figure (\ref{intbyparts}) we plotted 
\be{R}
R=\log_{10}\left|x^n\frac{d}{dx}\paq{2\,\bar x^n\pa{1+\frac{6}{\bar x^4}\frac{\M^2}{h^2}\frac{k^3}{\bar k^3}}}^{-1}\right|
\ee
i.e. the base 10 logarithm of the ratio between the modulus of the complex integrand of (\ref{rest}) and the modulus of the original integrand on the l.h.s. of (\ref{intbypart}). Such a ratio has been plotted for $n=3,4$ over a large $x$ interval and for different choices of $\frac{\M^2}{h^2}=10^4,10^7,10^{10},10^{13}$ (and $k/\bar k=1$) and is always much less then one, decreasing for increasing values of $\frac{\M^2}{h^2}$.\\
The factor $1+\frac{6}{x^4}\frac{\M^2}{h^2}\frac{k^3}{\bar k^3}$ is order one for $x^4\gtrsim \frac{\M^2}{h^2}\frac{k^3}{\bar k^3}\gg 1$ and, in particular is much grater than one for $x^4\ll \frac{\M^2}{h^2}\frac{k^3}{\bar k^3}$. Given our assumptions for $\bar k$ and $\M/h$ the contributions we are neglecting are further suppressed for $x$ large since the amplitude of the oscillation is inversely proportional to $x^3$, $x^4$. We then conclude that our estimate (\ref{intbypart}) is very robust.\\ 
For each $k$-mode the amplitude $c_2$ is then given by
\be{c2fin}
c_2^{(dS)}=\left.-\sqrt{2}\re^{2i\alpha_0}\frac{\pa{1-2i k \bar\eta}\re^{2i k \bar\eta+4i \frac{\M^2}{h^2}\frac{k^3}{\bar k^3}\frac{1}{(-k \bar\eta)^{3}}+i\frac{\pi}{2}}}{2(-k \bar \eta)^4\pa{1+\frac{6}{(-k \bar\eta)^4}\frac{\M^2}{h^2}\frac{k^3}{\bar k^3}}}\right|_{\eta_{0}}^{\eta}.
\ee
This result is the difference of an expression evaluated at $\bar\eta=\eta\rightarrow 0^{-}$ (and correspondingly $x\rightarrow 0^+$) and the same expression at $\bar \eta=\eta_0$ (and $\bar x=x_0$). In particular this latter $\eta_0$ can be the value of the conformal time at the moment when the transition between quantum and classical gravity begins and the QGE can be treated perturbatively (for example at the beginning of inflation \cite{BO-cosm0} or suffciently after the bounce). If we assume that at $\eta_0$ the perturbations are in the BD vacuum, well inside the horizon and in particular $x_0^4\gg \frac{\M^2}{h^2}\frac{k^3}{\bar k^3}$ then (\ref{c2fin}) becomes
\be{c2finSIM}
c_2^{(dS)}\simeq-i\frac{\sqrt{2}}{12}\frac{h^2}{\M^2}\frac{\bar k^3}{k^3}\re^{2i\alpha_0}
\paq{\re^{4i \frac{\M^2}{h^2}\frac{k^3}{\bar k^3}\frac{1}{(-k \eta)^{3}}}-12i\frac{\M^2}{h^2}\frac{\re^{2i k \eta_0}}{(-\bar k \eta_0)^3}}
\ee
\begin{widetext}
The modifications to the inflationary spectra are given by:
\ba{dek}
\Delta_k=-2\epsilon \,\frac{k\eta+i}{k\eta-i}\re^{2i\alpha_0}\frac{\pa{1-2i k \eta}\re^{4i \frac{\M^2}{h^2}\frac{k^3}{\bar k^3}\frac{1}{(-k \eta)^{3}}+i\frac{\pi}{2}}}{2(-k \eta)^4\pa{1+\frac{6}{(-k \eta)^4}\frac{\M^2}{h^2}\frac{k^3}{\bar k^3}}}+{\rm c.c.}
\ea
which, in the long wavelength limit $-k\eta\rightarrow 0$, becomes
\be{dekL}
\Delta_k^{L}\simeq -\epsilon\frac{i}{6}\frac{h^2}{\M^2}\frac{\bar k^3}{k^3}
\paq{\re^{4i \frac{\M^2}{h^2}\frac{k^3}{\bar k^3}\frac{1}{(-k \eta)^{3}}-2ik\eta_0}-\frac{\M^2}{h^2}\frac{12i}{(-\bar k \eta_0)^3}}+{\rm c.c.}
\ee
\end{widetext}
where the oscillations left have a quantum gravitational origin and in the long wavelength limit average to zero. The final correction is
\be{findek}
\Delta_k^{L}\simeq \epsilon
\frac{4}{(k \eta_0)^3}.
\ee
We observe that depending on the sign of epsilon it can lead either to an increase or a decrease of power for large scales. Its value presumably originates from the dynamics during the bouncing phase which is unknown.  We can finally evaluate the product $k\eta_0$. Such a product is much greater than one since its is well inside the horizon. If we take $\eta_0$ as the conformal time at the beginning of inflation then $h_0\sim h_*\sim 10^{-5}\M$.  Given our estimate (\ref{a0bounce}), when $h_B\sim 0\ll h_*$, then $a_0\gg a_B$ and 
\be{eta0est}
k\eta_0\equiv\frac{k}{a_0h_0}\ll \frac{k}{10\; \M^{-1}\cdot 10^{-5}\M}=10^4k
\ee
leading to a correction $k^3\Delta_k\gg10^{-12}$. Let us note that although the correction is small its functional dependence on $k$ looks interesting since other QGE obtained from the WdW equation have the same form. In \cite{K} the same dependence on $k$ emerges from the QGE associated with the quantum operators in the MS equation which lead to non adiabatic transitions. For this latter case, however, a definite sign in front of the corrections was obtained and the suppression of these effects was proportional to the ratio $h_*^2/\M^2$. Further the uncertainty related to the unknown $\bar k$ scale was present. In contrast, in the approach presented here, the dependence on $\bar k$ simplifies and, given the uncertainty associated with $\epsilon$ and $a_B$, we argue that the QGE may be larger then those calculated in \cite{K}.  

\section{Conclusions}
We studied QGE on the primordial spectra associated with the emergence of time for a quantum inflationary universe. Such effects are the consequence of the quantum behaviour of the gravitational sector (scale factor) and are usually ignored in the standard approaches to the study of the evolution of inflationary perturbations. The latter approaches rely on the assumption that both the homogeneous scale factor and inflatonic degrees of freedom follow a classical trajectory and the perturbations evolve quantum mechanically. The evolution of the perturbations then follows that of the homogeneous sector and a time parameter for them is usually defined in terms of the (quasi) classical trajectory of the scale factor. Our study, on the other hand, is based on the WdW canonical quantisation scheme of the inflaton-gravity system and takes into consideration the quantum effects both for the inflaton and the scale factor. In principle, without a classical limit for the scale factor even the definition of time appears difficult. In our previous article \cite{TVV} we already studied the problem of the emergence of time for the WdW approach to quantum cosmology and addressed the case of a highly quantum state for gravity. Correspondingly a highly peaked state for the homogeneous inflaton was considered.\\
The problem of the emergence of time is not a distinctive feature of quantum cosmology but generically emerges in closed quantum systems where the (quantum) evolution of a part of such systems must be parametrised by that of their remaining part which, usually, is supposed to evolve classically. If such is not the case time can still be introduced by means of the quantum probability which can be associated with the inverse of a ``velocity'' \cite{rowe} in analogy with its classical counterpart. In \cite{TVV} this latter approach was followed. Furthermore we eliminated the residual quantum effects by coarse graining over trans-planckian oscillations of the gravitational wave-function. Such a procedure is justified if the quantum degrees of freedom of which we must study the evolution have energies well below the Planck one. The quantum fluctuations enhanced by inflation have energies which are red-shifted of many order of magnitude and are certainly trans-Planckian at some time during their evolution. For these we expect that QGE may have some impact at some stage \cite{trans}.\\
In the context illustrated above the dynamics of inflationary perturbations is calculated and the QGE associated with the quantum behaviour of gravity are evaluated as perturbations to the standard evolution. Let us note that the magnitude of such effect is related to the initial conditions of the inflationary phase. In particular we restricted our analysis to the cases which can be studied by a perturbative expansion. For these the initial conditions involve a coherent superposition of an incoming and an outgoing wave and the (perturbatively small) incoming wave is originated by a bounce dynamics which emerges from the solution of the homogeneous WdW equation in the limit for $a$ small. Indeed in the presence of a reflecting barrier (or bounce) one can envisage the possibility of an interference between incoming and outgoing universes and a breakdown of time reversal invariance will appear in our expanding universe \cite{BO-cosm0}. Since the effect is larger for small $k$ this could be related to the observed power loss on large scales of the spectrum of the the temperature anisotropies in CMB \cite{powerloss1}. The QGE which are not associated with the emergence of time are intentionally neglected, but they were previously calculated in a series of articles \cite{K} (they are non leading in the Planck mass squared).\\
Our method is finally applied to the de Sitter case for simplicity, and we found corrections proportional to $k^{-3}$. This is a recurrent feature of the QGE calculated in this framework as the same $k$ dependence has been found in \cite{K} and other papers. In contrast with the results found in \cite{K} the QGE are not proportional to the ratio $h^2/\M^2\ll 1$, however they still are small. Better estimates of their magnitude and their overall sign must be derived from the solution of the homogeneous WdW equation or, at least, from an improved description of the bouncing phase (and the matching between small and large $a$ solutions). If more accurate solutions were known, higher order effects (slow roll) could be added and more realistic scenario studied in more detail, we hope to return to this.

\section*{Acknowledgments}
The work of A.K. was partially supported by the RFBR grant No 17-02-01008. 


\end{document}